\documentclass[aps,preprint]{revtex4}%
\usepackage{amsfonts}
\usepackage{amsmath}
\usepackage{amssymb}
\usepackage{graphicx}%
\providecommand{\U}[1]{\protect \rule{.1in}{.1in}}

\begin{document}
\title{Bipolarons and multi-polarons consisting of impurity atoms in a Bose-Einstein condensate}
\author{W. Casteels$^{1}$, J. Tempere$^{1,2}$ and J. T. Devreese$^{1}$}
\affiliation{$^{1}$TQC, Universiteit Antwerpen, Universiteitsplein 1, 2610 Wilrijk, Belgium}
\affiliation{$^{2}$Lyman Laboratory of Physics, Harvard University, Cambridge,
Massachusetts 02138, USA}

\begin{abstract}
The variational Feynman formalism for the polaron, extended to an all-coupling
treatment of bipolarons, is applied for two impurity atoms in a Bose-Einstein
condensate. This shows that if the polaronic coupling strength is large enough
the impurities will form a bound state (the bipolaron). As a function of the
mutual repulsion between the impurities two types of bipolaron are
distinguished: a tightly bound bipolaron at weak repulsion and a dumbbell
bipolaron at strong repulsion. Apart from the binding energy, also the
evolution of the bipolaron radius and its effective mass are examined as a
function of the strength of the repulsive interaction between the impurities
and of the polaronic cupling strength. We then apply the strong-coupling
formalism to multiple impuritiy atoms in a condensate which leads to the
prediction of multi-polaron formation in the strong coupling regime. The
results of the two formalisms are compared for two impurities in a condensate
which results in a general qualitative agreement and a quantitative agreement
at strong coupling. Typically the system of impurity atoms in a Bose-Einstein
condensate is expected to exhibit the polaronic weak coupling regime. However,
the polaronic coupling strength is in principle tunable with a Feshbach resonance.

\end{abstract}
\maketitle

\section{Introduction}

In recent years systems related to ultracold gases have successfully been
aplied as quantum simulators for many-body theories \cite{RevModPhys.80.885}.
A specific example is a Bose-Einstein condensate (BEC) with an impurity atom
of which the Hamiltonian can be mapped onto the Fr\"{o}hlich polaron
Hamiltonian, provided the Bogoliubov approximation is valid
\cite{PhysRevB.46.301, PhysRevLett.96.210401, PhysRevA.73.063604}. The polaron
is a well-known concept in solid state physics where it represents the
quasiparticle that consisists of an electron in a polar or ionic lattice,
dressed with the self-induced polarization cloud which is described by the
lattice vibrations or phonons \cite{Devreese}. In the context of ultracold
atomic gases the electron is replaced by an impurity atom and the role of the
phonons is played by the Bogoliubov excitations of the condensate. Hitherto,
the Fr\"{o}hlich Hamiltonian has resisted analytical diagonalization, making
it the subject of various approximation schemes \cite{2010arXiv1012.4576D}. A
polaronic system typically exhibits different coupling regimes characterized
by a quasi-free polaron at weak coupling and a bound state in the self-induced
potential at strong coupling. The variational all-coupling treatment, as
developed by Feynman to describe the ground state of the polaron
\cite{PhysRev.97.660}, reveals the transition between the coupling regimes and
nicely interpolates between the Fr\"{o}hlich perturbative result at weak
coupling \cite{doi:10.1080/00018735400101213}\ and the Landau-Pekar
strong-coupling result \cite{LandauPekar1, LandauPekar}. For a single impurity
in a Bose-Einstein condensate the application of the Feynman all-coupling
theory also revealed the transition between the different coupling regimes
\cite{PhysRevB.80.184504, 0953-4075-43-10-105301}.

Recently there have been reports on the immersion of a single impurity in an
ultracold gas \cite{PhysRevLett.109.235301, PhysRevLett.109.123201,
citeulike:11199567, Widera1, PhysRevA.85.023623}. For neutral impurity atoms
in a BEC the system is expected to be in the weak polaronic coupling regime.
For a Li-6 impurity in a Na condensate for example the polaronic coupling
strength $\alpha$ is of the order of $10^{-3}$ \cite{PhysRevB.80.184504},
which is well within the weak coupling regime. However, the polaronic coupling
strength is in principle externally tunable and can be increased by means of a
Feshbach resonance. The feasibility of interspecies Feshbach resonances have
recently been demonstrated in various ultracold mixtures (see for example
Refs. \cite{PhysRevA.85.032506, PhysRevA.85.051602, PhysRevA.85.042721,
PhysRevA.87.010701}). An experimental study of the BEC-impurity polaron with a
variable coupling strength could shed light on discrepancies between different
predictions of the polaronic dynamic response properties
\cite{PhysRevA.83.033631, PhysRevLett.91.236401, PhysRevB.5.2367}.

Considering multiple impurities in a Bose-Einstein condensate unveils a whole
range of interesting phenomena. The presence of the Bose-Einstein condensate
induces an effective interaction between the impurities
\cite{PhysRevA.61.053605, PhysRevA.61.053601}. At weak polaronic coupling the
application of the many-polaron formalism, as developed in
Refs.\  \cite{PSSB:PSSB2220820204, PhysRevB.64.104504}, leads to a description
of the ground state properties and the response to Bragg spectroscopy of
ultracold weakly interacting binary mixtures \cite{PhysRevA.84.063612}. If the
polaronic coupling is strong enough the BEC-induced interaction can lead to
the clustering of the impurities, also known as a multi-polaron
\cite{1367-2630-13-10-103029, PhysRevLett.102.025301}. The possibility of the
formation of a multi-polaron at strong polaronic coupling is also well-known
in the solid state context \cite{MultiPolaron1, Multipolaron2,
PhysRevB.47.2596, PhysRevB.71.214301, PhysRevLett.104.210402}. A special case
is the formation of a bound state of two electrons, commonly known as the
bipolaron \cite{0034-4885-72-6-066501, 0034-4885-57-12-001, PhysRevB.43.5296}.
Bipolarons have attracted much attention because of their possible role as an
unconventional pairing mechanism for high-temperature superconductivity
\cite{PhysRevB.77.094502, Alexandrov}.

In this work, we consider a few impurities in a Bose-Einstein condensate and
examine the formation of a multi-polaron. We start in section
\ref{section: mapping}\ by showing how the Hamtiltonian of impurity atoms in a
BEC can be mapped onto the Fr\"{o}hlich Hamiltonian, provided the Bogoliubov
approximation is valid. Then, in section \ref{section: JenFey}, the
all-coupling variational path integral approach, as developed by Feynman in
Ref. \cite{PhysRev.97.660} for a single polaron and extended in Ref.
\cite{PhysRevB.43.2712} for two electrons in a polar or ionic lattice, is
applied to two distinguishable impurities in a condensate. This allows us to
examine the polaron-bipolaron transition and the bipolaron ground state
properties such as the radius and the effective mass. In section
\ref{section: SC} a strong-coupling treatment for multiple impurities in a
BEC, based on the Landau-Pekar strong-coupling approach, is considered. This
is first applied for two impurities in a condensate to examine the
polaron-bipolaron transition and the polaronic properties, and then to
multiple impurities to study the formation of a larger multi-polaron. Finally,
in section \ref{section: comp}, we compare the results of the two presented
formalisms for two distinguishable impurities in a BEC and in section
\ref{section: concl} we conclude.

\section{The polaronic system consisting of multiple imurities in a
Bose-Einstein condensate\label{section: mapping}}

The Hamiltonian of $N_{I}$ impurity atoms, in the presence of a homogeneous
Bose gas is given by:%
\begin{align}
\widehat{H}  &  =\sum_{i=1}^{N_{I}}\frac{\widehat{p}_{i}^{2}}{2m_{I}}%
+\sum_{\vec{k}}E_{\vec{k}}\widehat{a}_{\vec{k}}^{\dag}\widehat{a}_{\vec{k}%
}+\frac{1}{2}\sum_{\vec{k},\vec{k}^{\prime},\vec{q}}V_{BB}\left(  \vec
{q}\right)  \widehat{a}_{\vec{k}^{\prime}-\vec{q}}^{\dag}\widehat{a}_{\vec
{k}+\vec{q}}^{\dag}\widehat{a}_{\vec{k}}\widehat{a}_{\vec{k}^{\prime}}%
+\sum_{\vec{k},\vec{q}}V_{IB}\left(  \vec{q}\right)  \widehat{\rho}_{I}\left(
\vec{q}\right)  \widehat{a}_{\vec{k}-\vec{q}}^{\dag}\widehat{a}_{\vec{k}%
}\nonumber \\
&  +\sum_{i<j}^{N_{I}}V_{II}\left(  \widehat{\vec{r}}_{i}-\widehat{\vec{r}%
}_{j}\right)  . \label{Ham1}%
\end{align}
The first term represents the kinetic energy of the impurities of mass $m_{I}$
and associated momentum operators $\left \{  \widehat{\vec{p}}_{i}\right \}  $
and position operators $\left \{  \widehat{\vec{r}}_{i}\right \}  $. The
operators $\widehat{a}_{\vec{k}}^{\dag}$, $\widehat{a}_{\vec{k}}$ create,
respectively annihilate, a boson with mass $m_{B}$, wave vector $\vec{k}$ and
energy $E_{\vec{k}}=\left(  \hbar k\right)  ^{2}/2m_{B}-\mu$, with $\mu$ the
boson chemical potential. These bosons interact mutually, with $V_{BB}\left(
\vec{q}\right)  $ the Fourier transform of the interaction potential. The
Fourier transform of the impurity-boson interaction potential is
$V_{IB}\left(  \vec{q}\right)  $ and couples the boson density to the impurity
density
\begin{equation}
\widehat{\rho}_{I}\left(  \vec{q}\right)  =\sum_{i=1}^{N_{I}}e^{i\vec
{q}.\widehat{\vec{r}}_{i}}.
\end{equation}
The impurity-impurity interaction is described by the interaction potential
$V_{II}\left(  \vec{r}\right)  $. For the interparticle interactions we assume
contact pseudo potentials: $V_{BB}\left(  \vec{r}\right)  =g_{BB}\delta \left(
\vec{r}\right)  $, $V_{IB}\left(  \vec{r}\right)  =g_{IB}\delta \left(  \vec
{r}\right)  $ and $V_{II}\left(  \vec{r}\right)  =g_{II}\delta \left(  \vec
{r}\right)  $. The interaction strengths $g_{BB}$, $g_{IB}$ and $g_{II}$ are
related to the corresponding scatttering lengths $a_{BB}$, $a_{IB}$ and
$a_{II}$ through the Lippmann-Schwinger equation. For the boson-boson and the
impurity-impurity interactions the first order result suffices: $g_{BB}%
=4\pi \hbar^{2}a_{BB}/m_{B}$ and $g_{II}=4\pi \hbar^{2}a_{II}/m_{I}$. We will
only consider repulsive impurities ($a_{II}>0$). For the impurity-boson
interaction the Lippmann-Schwinger equation has to be treated up to second
order to obtain convergent final results:%
\begin{equation}
\frac{2\pi \hbar^{2}a_{IB}}{m_{r}}=g_{IB}-g_{IB}^{2}\sum_{\vec{k}\neq0}%
\frac{2m_{r}}{\left(  \hbar k\right)  ^{2}}, \label{LippSchwingSecOr}%
\end{equation}
where $m_{r}$ is the reduced mass ($m_{r}^{-1}=m_{I}^{-1}+m_{B}^{-1}$).

If a Bose-Einstein condensate is realized the number of bosons $N_{0}$\ that
occupy the single-particle ground state becomes a macroscopic number:
$N_{0}\gg1$ \cite{PhysRev.104.576}. This is expressed by the Bogoliubov shift
\cite{BogolShift}, which transforms the Hamiltonian (\ref{Ham1}) into:
\begin{equation}
\widehat{H}=E_{GP}+N_{I}N_{0}g_{IB}+\widehat{H}_{pol}^{\left(  N_{I}\right)
}. \label{HamTot}%
\end{equation}
Where $E_{GP}$ is the Gross-Pitaevskii energy of the homogeneous condensate
\cite{Pitaev, Gross}. The second term in the Hamiltonian (\ref{HamTot}) is the
interaction shift due to the impurities and the third term is the Fr\"{o}hlich
Hamiltonian for $N_{I}$ polarons:%
\begin{equation}
\widehat{H}_{pol}^{\left(  N_{I}\right)  }=\sum_{i=1}^{N_{I}}\frac{\widehat
{p}_{i}^{2}}{2m_{I}}+\sum_{\vec{k}}\hbar \omega_{\vec{k}}\widehat{\alpha}%
_{\vec{k}}^{\dag}\widehat{\alpha}_{\vec{k}}+\sum_{\vec{k}}V_{\vec{k}}%
\widehat{\rho}_{I}\left(  \vec{q}\right)  \left(  \widehat{\alpha}_{\vec{k}%
}+\widehat{\alpha}_{-\vec{k}}^{\dag}\right)  +\sum_{i<j}^{N_{I}}V_{II}\left(
\widehat{\vec{r}}_{i}-\widehat{\vec{r}}_{j}\right)  . \label{HamNiPol}%
\end{equation}
The first term of the Fr\"{o}hlich Hamiltonian represents the kinetic energy
of the impurities. The operators $\widehat{\alpha}_{\vec{k}}^{\dag}$,
$\widehat{\alpha}_{\vec{k}}$ create, respectively annihilate, a Bogoliubov
excitation with wave number $\vec{k}$ and energy%
\begin{equation}
\hbar \omega_{\vec{k}}=\frac{\hbar^{2}k}{2m_{B}\xi}\sqrt{2+\left(  \xi
k\right)  ^{2}}. \label{BogDisp}%
\end{equation}
Where the healing length of the Bose condensate was introduced: $\xi
=1/\sqrt{8\pi a_{BB}n_{0}}$, with $n_{0}=N_{0}/V$ the condensate density (we
work with unit volume for the homogenous gas). The third term in the
Fr\"{o}hlich Hamiltonian (\ref{HamNiPol}) describes the interaction between
the impurities and the Bogoliubov excitations with the interaction amplitude%
\begin{equation}
V_{\vec{k}}=\sqrt{N_{0}}g_{IB}\left[  \frac{\left(  \xi k\right)  ^{2}%
}{\left(  \xi k\right)  ^{2}+2}\right]  ^{1/4}. \label{IntAmp}%
\end{equation}
The last term of the Fr\"{o}hlich Hamiltonian (\ref{HamNiPol}) corresponds to
the interaction between the impurities. The Fr\"{o}hlich Hamiltonian
(\ref{HamNiPol}) was originally derived to describe the interaction between an
electron (or hole) and the longitudinal optical phonons in a polar or ionic
crystal \cite{doi:10.1080/00018735400101213}. A strong repulsion beween the
impurity and the bosons can lead to a large depletion of the condensate in the
vicinity of the impurity which can break the validity of the Bogoliubov
approximation and the applicability of the Frohlich Hamiltonian to describe
the system. This typically leads to a bubble state \cite{2013arXiv1304.7704B}.

\section{All-coupling variational treatment for two impuritiy atoms in a
BEC\label{section: JenFey}}

\subsection{All-coupling formalism}

\begin{figure}
[ptb]
\begin{center}
\includegraphics[
height=3.3411cm,
width=12.1166cm
]%
{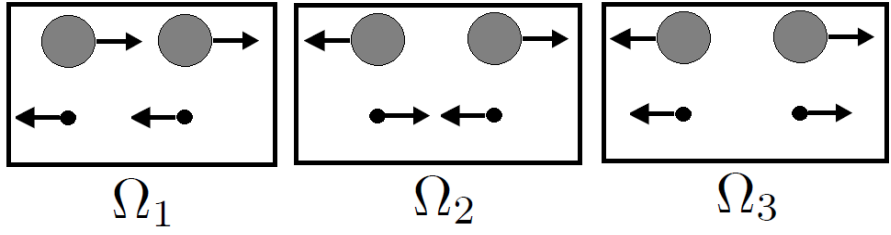}%
\caption{Schematic presentation of the eigenmodes corresponding to the
eigenfrequencies $\Omega_{1}$, $\Omega_{2}$ and $\Omega_{3}$. The small black
dots represent the impurities and the larger grey dots depict the fictitious
particles of mass $M$.}%
\label{fig: eigenmodes}%
\end{center}
\end{figure}

We consider the generic polaronic system of two distinguishable particles
interacting with a bosonic bath through the Fr\"{o}hlich interaction, i.e. any
system that can be described by the Hamiltonian (\ref{HamNiPol}) with
$N_{I}=2$. The variational all-coupling single polaron treatment, as
originally developed by Feynman \cite{PhysRev.97.660}, was extended in Ref.
\cite{PhysRevB.43.2712}\ to the case of two polaronic particles. This approach
is based on the Jensen-Feynman variational inequality for the free energy
$\mathcal{F}$ \cite{Feynman, Kleinert}:
\begin{equation}
\mathcal{F}\leq \mathcal{F}_{0}+\frac{1}{\hbar \beta}\left \langle \mathcal{S}%
-\mathcal{S}_{0}\right \rangle _{\mathcal{S}_{0}}, \label{JensFeyn}%
\end{equation}
with $\mathcal{S}$ the action of the polaronic system as described by the
Fr\"{o}hlich Hamiltonian (\ref{HamNiPol}), $\mathcal{S}_{0}$ the action of a
variational trial system with free energy $\mathcal{F}_{0}$ and $\beta$ the
inverse temperature $T$: $\beta=\left(  k_{B}T\right)  ^{-1}$. Eliminating the
degrees of freedom of the Bogoliubov excitations leads to an effective polaron
action, containing retardation effects:%
\begin{align}
\mathcal{S}  &  \mathcal{=}\int_{0}^{\hbar \beta}d\tau \left[  \sum_{i=1}%
^{N_{I}}\frac{m_{I}}{2}\dot{r}_{i}^{2}\left(  \tau \right)  +\sum_{i<j}%
V_{II}\left(  \vec{r}_{i}-\vec{r}_{j}\right)  \right] \nonumber \\
&  -\sum_{j,l=1}^{N_{I}}\sum_{\vec{k}}\frac{\left \vert V_{\vec{k}}\right \vert
^{2}}{\hbar}\int_{0}^{\hbar \beta}d\tau \int_{0}^{\hbar \beta}d\sigma
G_{\omega_{\vec{k}}}\left(  \tau-\sigma \right)  e^{i\vec{k}.\left[  \vec
{r}_{j}\left(  \tau \right)  -\vec{r}_{l}\left(  \sigma \right)  \right]  },
\end{align}
with $G_{\omega_{\vec{k}}}\left(  u\right)  $ the Green's function of the
Bogoliubov excitations:%
\begin{equation}
G_{\omega_{\vec{k}}}\left(  u\right)  =\frac{\cosh \left[  \omega_{\vec{k}%
}\left(  \hbar \beta/2-\left \vert u\right \vert \right)  \right]  }%
{2\sinh \left(  \beta \hbar \omega_{\vec{k}}/2\right)  }.
\end{equation}
For a single polaron a variational system was suggested by Feynman that mimics
the influence of the interaction with the Bogoliubov excitations on the
impurity by a harmonic coupling to a fictitious particle of mass $M$ with
oscillator strength $\kappa$ \cite{PhysRev.97.660}. The upper bound for the
free energy (\ref{JensFeyn})\ is then minimized as a function of the
variational parameters $M$ and $\kappa$. In Ref. \cite{PhysRevB.43.2712} an
extension of this trial system was introduced for the case of two polaronic
particles which is schematically presented in Fig. \ref{fig: modelsysteem}. As
is the case in the Feynman one-polaron trial system the impurities interact
quadratically with fictitious particles of mass $M$ with oscillator strength
$\kappa$. Furtermore, there is a quadratic interaction, with oscillator
strength $\kappa^{\prime}$, with the fictitious particle of the other
impurity. The particles are separated by the vector $\vec{a}$ and they
mutually interact quadratically with strength $K$. For $\kappa^{\prime}=K=0$
this reduces to (twice) the Feynman model system. After transforming to the
eigenmodes the action of the trial system can be written as:%

\begin{figure}
[ptb]
\begin{center}
\includegraphics[
height=2.4154in,
width=2.0384in
]%
{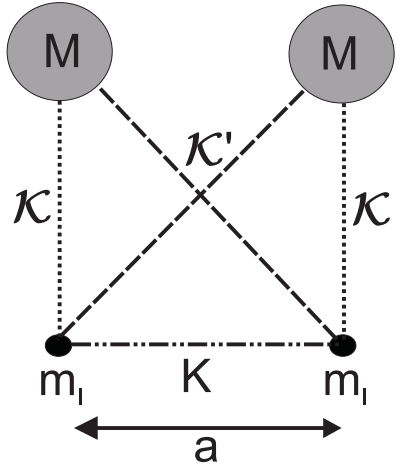}%
\caption{Schematical picture of the variational trial system for two polaronic
particles, as introduced in Ref. \cite{PhysRevB.43.2712}. The black dots
represent the impurity atoms of mass $m_{I}$ and the larger grey dots depict
the fictitious particles of mass $M$. The connecting lines represent harmonic
interactions with the corresponding oscillator strengths indicated.}%
\label{fig: modelsysteem}%
\end{center}
\end{figure}
\begin{equation}
\mathcal{S}_{0}\mathcal{=}\int_{0}^{\hbar \beta}d\tau \left[  \frac{\mu_{0}}%
{2}\dot{\rho}_{0}^{2}+\sum_{j=1}^{3}\left(  \frac{\mu_{j}}{2}\dot{\rho}%
_{j}^{2}+\frac{1}{2}\mu_{j}\Omega_{j}^{2}\rho_{j}^{2}\right)  \right]  ,
\label{Actietrial}%
\end{equation}
with $\left \{  \vec{\rho}_{i}\right \}  $ the coordinates of the eigenmodes of
the trial system and $\left \{  \Omega_{i}\right \}  $ the corresponding
eigenfrequencies ($\Omega_{0}=0$ corresponds to a translation of the trial
system as a whole):%
\begin{align}
\Omega_{1}^{2}  &  =\frac{M+m_{I}}{Mm_{I}}\left(  \kappa+\kappa^{\prime
}\right)  ;\\
\Omega_{2,3}^{2}  &  =\frac{1}{2}\left[  \frac{M+m_{I}}{Mm_{I}}\left(
\kappa+\kappa^{\prime}\right)  -\frac{2K}{m_{I}}\pm \sqrt{\left[  \frac
{M-m_{I}}{Mm_{I}}\left(  \kappa+\kappa^{\prime}\right)  -\frac{2K}{m_{I}%
}\right]  ^{2}+\frac{4}{m_{I}M}\left(  \kappa-\kappa^{\prime}\right)  ^{2}%
}\right]  .
\end{align}
Since all oscilator strengths are positive the eigenfrequencies satisfy the
inequalities%
\begin{align}
\Omega_{1}^{2}  &  \geq \Omega_{2}^{2}+\Omega_{3}^{2};\\
\Omega_{2}  &  \geq \nu \geq \Omega_{3}\geq0.
\end{align}
Where we introduced the frequency parameter $\nu=\sqrt{\left(  \kappa
+\kappa^{\prime}\right)  /M}$. In expression (\ref{Actietrial}) the following
mass factors were introduced:%
\begin{equation}
\mu_{0}=2\left(  m_{I}+M\right)  ;\text{ \ }\mu_{2}=\frac{2m_{I}M}{\left(
m_{I}+M\right)  };\text{ \ }\mu_{2}=1;\text{ \ }\mu_{3}=1.
\end{equation}
The corresponding eigenmodes of the trial system are schematically presented
in Fig. \ref{fig: eigenmodes}. The action (\ref{Actietrial}) shows that the
trial system decouples in a free particle and three harmonic oscillators.%

After eliminating the degrees of freedom of the two fictitious particles of
mass $M$ the effective action of the trial system becomes retarded and can be
written as%
\begin{align}
\mathcal{S}_{0}  &  =\int_{0}^{\hbar \beta}d\tau \left[  \sum_{i=1,2}\frac
{m_{I}}{2}\dot{r}_{i}^{2}\left(  \tau \right)  -\frac{K}{2}\left(  \vec{r}%
_{1}-\vec{r}_{2}-\vec{a}\right)  ^{2}\right] \nonumber \\
&  +\frac{\kappa^{2}+\kappa^{\prime2}}{4M\nu}\int_{0}^{\hbar \beta}d\tau
\int_{0}^{\hbar \beta}d\sigma G_{\nu}\left(  \tau-\sigma \right)  \sum
_{j}\left[  \vec{r}_{j}\left(  \tau \right)  -\vec{r}_{j}\left(  \sigma \right)
\right]  ^{2}\nonumber \\
&  +\frac{\kappa \kappa^{\prime}}{M\nu}\int_{0}^{\hbar \beta}d\tau \int
_{0}^{\hbar \beta}d\sigma G_{\nu}\left(  \tau-\sigma \right)  \left[  \vec
{r}_{1}\left(  \tau \right)  -\vec{r}_{2}\left(  \sigma \right)  -\vec
{a}\right]  ^{2}.
\end{align}
Applying the Jensen-Feynman inequality then results in an upper bound $E$ for
the polaronic contribution to the ground state energy at temperature zero
\cite{PhysRevB.43.2712}:%
\begin{align}
E  &  =\sum_{j=1}^{3}\frac{3}{2}\hbar \Omega_{j}-3\hbar \nu+\sum_{\vec{k}}%
V_{II}\left(  \vec{k}\right)  e^{i\vec{k}.\vec{a}}e^{-k^{2}D_{12}\left(
0\right)  }\nonumber \\
&  -2\sum_{\vec{k}}\frac{\left \vert V_{\vec{k}}\right \vert ^{2}}{\hbar}%
\int_{0}^{\infty}due^{-\omega_{\vec{k}}u}\left[  e^{-k^{2}D_{11}\left(
u\right)  }+e^{i\vec{k}.\vec{a}}e^{-k^{2}D_{12}\left(  u\right)  }\right]
\nonumber \\
&  -\frac{3}{2}\frac{\Omega_{1}^{2}-\nu^{2}}{\Omega_{1}^{2}}\frac{\hbar
\Omega_{1}}{2}-\frac{3}{2}\frac{\Omega_{2}^{2}-\nu^{2}}{\Omega_{2}^{2}%
-\Omega_{3}^{2}}\frac{\hbar \Omega_{2}}{2}-\frac{3}{2}\frac{\nu^{2}-\Omega
_{3}^{2}}{\Omega_{2}^{2}-\Omega_{3}^{2}}\frac{\hbar \Omega_{3}}{2}.
\label{UpperBound}%
\end{align}
Where the functions $D_{11}\left(  u\right)  $ and $D_{12}\left(  u\right)  $
are defined as:%
\begin{align}
D_{11}\left(  u\right)   &  =\frac{\hbar}{2m_{I}}\left[  \frac{\nu^{2}}%
{\Omega_{1}^{2}}\frac{u}{2}+\frac{\Omega_{1}^{2}-\nu^{2}}{\Omega_{1}^{2}%
}E\left(  \Omega_{1},u\right)  +\frac{\Omega_{2}^{2}-\nu^{2}}{\Omega_{2}%
^{2}-\Omega_{3}^{2}}E\left(  \Omega_{2},u\right)  +\frac{\nu^{2}-\Omega
_{3}^{2}}{\Omega_{2}^{2}-\Omega_{3}^{2}}E\left(  \Omega_{3},u\right)  \right]
;\label{D11Fun}\\
D_{12}\left(  u\right)   &  =\frac{\hbar}{2m_{I}}\left[  \frac{\nu^{2}}%
{\Omega_{1}^{2}}\frac{u}{2}+\frac{\Omega_{1}^{2}-\nu^{2}}{\Omega_{1}^{2}%
}E\left(  \Omega_{1},u\right)  +\frac{\Omega_{2}^{2}-\nu^{2}}{\Omega_{2}%
^{2}-\Omega_{3}^{2}}F\left(  \Omega_{2},u\right)  +\frac{\nu^{2}-\Omega
_{3}^{2}}{\Omega_{2}^{2}-\Omega_{3}^{2}}F\left(  \Omega_{3},u\right)  \right]
; \label{D12Fun}%
\end{align}
with:%
\begin{align}
E\left(  \Omega,u\right)   &  =\frac{1-\exp \left[  -\Omega u\right]  }%
{2\Omega};\\
F\left(  \Omega,u\right)   &  =\frac{1+\exp \left[  -\Omega u\right]  }%
{2\Omega}.
\end{align}
The next step is to minimize the upper bound $E$ (\ref{UpperBound}) as a
function of the variational parameters $\left \{  \nu,\Omega_{1},\Omega
_{2},\Omega_{3},\vec{a}\right \}  $.

\subsubsection{Bipolaron radius}

As an estimate for the bipolaron radius $R$\ the root mean square of the
distance between the impurities is used \cite{PhysRevB.43.2712}:%
\begin{align}
R  &  =\sqrt{\left \langle \left[  \vec{r}_{1}\left(  \tau \right)  -\vec{r}%
_{2}\left(  \tau \right)  \right]  ^{2}\right \rangle }\nonumber \\
&  =\sqrt{a^{2}+6D_{12}\left(  0\right)  }. \label{BipRadius}%
\end{align}
whith the function $D_{12}\left(  u\right)  $ as defined in (\ref{D12Fun}).

\subsubsection{Effective mass of the bipolaron}

The effective mass $m^{\ast}$ can be derived from the path integral
propagation from $\vec{r}_{i}\left(  0\right)  $ to $\vec{r}_{i}\left(
T\right)  =\vec{r}_{i}\left(  0\right)  +\vec{U}T$, for $i=1,2$. The ground
state energy then behaves as%
\begin{equation}
E\left(  U\right)  =E\left(  0\right)  +\frac{m^{\ast}U^{2}}{2}.
\end{equation}
This procedure was implemented by Feynman to derive the effective mass of a
single polaron at arbitrary coupling \cite{PhysRev.97.660}. The same treatment
for two particles leads to an expression for the effective mass of the
bipolaron:%
\begin{equation}
m^{\ast}=2m_{I}+2\sum_{\vec{k}}\frac{\left \vert V_{\vec{k}}\right \vert ^{2}%
}{\hbar}\int_{0}^{\infty}due^{-\omega_{\vec{k}}u}\left[  e^{-k^{2}%
D_{11}\left(  u\right)  }+e^{i\vec{k}.\vec{a}}e^{-k^{2}D_{12}\left(  u\right)
}\right]  u^{2}k_{z}^{2}. \label{EffMassBip}%
\end{equation}
With the functions $D_{11}\left(  u\right)  $ and $D_{12}\left(  u\right)  $
as defined in (\ref{D11Fun}) and (\ref{D12Fun}), respectively.

\subsubsection{Single polaron limit}

The trial system reduces to (twice) the Feynman one-polaron trial system for
$K=\kappa^{\prime}=0$. For the eigenmodes this corresponds to $\Omega_{3}=0$
and $\Omega_{1}=\Omega_{2}=\Omega$. This gives for the functions
$D_{11}\left(  u\right)  $ (\ref{D11Fun}) and $D_{12}\left(  u\right)  $
(\ref{D12Fun}):%
\begin{align}
\lim_{\substack{\Omega_{3}\rightarrow0\\ \Omega_{1}=\Omega_{2}=\Omega}%
}D_{11}\left(  u\right)   &  =\frac{\hbar}{2m_{I}}\left[  \frac{\nu^{2}%
}{\Omega^{2}}u+\frac{\Omega^{2}-\nu^{2}}{\Omega^{2}}\frac{1-\exp \left[
-\Omega u\right]  }{\Omega}\right]  =D\left(  u\right)  ;\\
\lim_{\substack{\Omega_{3}\rightarrow0\\ \Omega_{1}=\Omega_{2}=\Omega}%
}D_{12}\left(  u\right)   &  \rightarrow \infty.
\end{align}
The upper bound for the ground state polaron energy (\ref{UpperBound}) becomes
in this limit:%
\begin{equation}
\lim_{\substack{\Omega_{3}\rightarrow0\\ \Omega_{1}=\Omega_{2}=\Omega
}}E=2\left[  \frac{3}{2}\hbar \left(  \Omega-\nu \right)  -\frac{3}{4}%
\frac{\Omega^{2}-\nu^{2}}{\Omega^{2}}\hbar \Omega-\sum_{\vec{k}}\frac
{\left \vert V_{\vec{k}}\right \vert ^{2}}{\hbar}\int_{0}^{\infty}%
due^{-\omega_{\vec{k}}u}e^{-k^{2}D\left(  u\right)  }\right]  .
\end{equation}
This is (twice) the upper bound for the ground state of a single polaron, as
derived by Feynman \cite{PhysRev.97.660}. The effective mass of the bipolaron
becomes in this limit:%
\begin{equation}
\lim_{\substack{\Omega_{3}\rightarrow0\\ \Omega_{1}=\Omega_{2}=\Omega}}m^{\ast
}=2\left(  m_{I}+\frac{1}{3}\sum_{\vec{k}}\frac{\left \vert V_{\vec{k}%
}\right \vert ^{2}}{\hbar}\int_{0}^{\infty}due^{-\omega_{\vec{k}}u}%
e^{-k^{2}D\left(  u\right)  }u^{2}k^{2}\right)  ,
\end{equation}
which is (twice) the single polaron effective mass, as derived by Feynman
\cite{PhysRev.97.660}.

\subsection{Two impurities in a Bose-Einstein condensate}

We now consider the specific system of two impurity atoms in a Bose-Einstein
condensate. Using the contact pseudo potential with the first order
Lippmann-Schwinger result gives for the impurity-impurity interaction:%
\begin{equation}
\sum_{\vec{k}}V_{II}\left(  \vec{k}\right)  e^{i\vec{k}.\vec{a}}%
e^{-k^{2}D_{12}\left(  0\right)  }=\frac{\hbar^{2}a_{II}}{2\sqrt{\pi}%
m_{I}D_{12}\left(  0\right)  ^{3/2}}e^{-\frac{a^{2}}{4D_{12}\left(  0\right)
}}.
\end{equation}
Introducing the interaction amplitude (\ref{IntAmp}) gives for the upper bound
for the ground state energy (\ref{UpperBound}) in polaronic units
($\hbar=m_{I}=\xi=1$):%
\begin{align}
E  &  =\sum_{j=1}^{3}\frac{3}{2}\Omega_{j}-3\nu+\frac{a_{II}}{2\sqrt{\pi
}D_{12}\left(  0\right)  ^{3/2}}e^{-\frac{a^{2}}{4D_{12}\left(  0\right)  }%
}\nonumber \\
&  +\frac{\alpha}{2\pi}\left(  \frac{m_{B}+1}{m_{B}}\right)  ^{2}\int
_{0}^{\infty}dk\left \{  \frac{2m_{B}}{m_{B}+1}\right. \nonumber \\
&  \left.  -\frac{k^{3}}{\sqrt{k^{2}+2}}\int_{0}^{\infty}due^{-\omega_{\vec
{k}}u}\left[  e^{-k^{2}D_{11}\left(  u\right)  }+\frac{\sin \left[  ak\right]
}{ak}e^{-k^{2}D_{12}\left(  u\right)  }\right]  \right \} \nonumber \\
&  -\frac{3}{2}\frac{\Omega_{1}^{2}-\nu^{2}}{\Omega_{1}^{2}}\frac{\Omega_{1}%
}{2}-\frac{3}{2}\frac{\Omega_{2}^{2}-\nu^{2}}{\Omega_{2}^{2}-\Omega_{3}^{2}%
}\frac{\Omega_{2}}{2}-\frac{3}{2}\frac{\nu^{2}-\Omega_{3}^{2}}{\Omega_{2}%
^{2}-\Omega_{3}^{2}}\frac{\Omega_{3}}{2}. \label{JensFeynTemp0}%
\end{align}
Where $\alpha$ is the dimensionless polaronic coupling parameter:%
\begin{equation}
\alpha=\frac{a_{IB}^{2}}{\xi a_{BB}}. \label{CouplPar}%
\end{equation}
The first term in the $k$-integrand in the right-hand side of expression
(\ref{JensFeynTemp0}) is a consequence of using the Lippmann-Schwinger
equation up to second order for the impurity-boson interaction strength
(\ref{LippSchwingSecOr}) in the second term of the total Hamiltonian
(\ref{HamTot}) and is needed for convergence. A similar procedure was applied
for the single-polaron all-coupling treatment in Ref.
\cite{PhysRevB.80.184504}. With the interaction amplitude (\ref{IntAmp}) the
bipolaron effective mass in polaronic units becomes%

\begin{equation}
m^{\ast}=2+\frac{\alpha}{8\pi^{2}}\left(  \frac{1+m_{B}}{m_{B}}\right)
^{2}\int d\vec{k}\sqrt{\frac{k^{2}}{k^{2}+2}}\int_{0}^{\infty}due^{-\omega
_{\vec{k}}u}\left[  e^{-k^{2}D_{11}\left(  u\right)  }+e^{i\vec{k}.\vec{a}%
}e^{-k^{2}D_{12}\left(  u\right)  }\right]  u^{2}k_{z}^{2}. \label{BipEffMass}%
\end{equation}

\subsection{Results and discussion}

For numerical calculations it is favorable to introduce a cutoff\ $K_{c}$ for
the $k$-integral in (\ref{JensFeynTemp0}). Similar to the one-polaron case we
use the inverse of the Van der Waals radius of the impurity-boson interaction
potential for $K_{c}$ \cite{PhysRevB.80.184504}. We introduce the specific
system of Li-6 impurities in a Na condensate which amounts to $m_{B}%
/m_{I}=3.8227$ and $\xi K_{c}=200$. The considered system with two
distinguishable impurities can for example be realized with two different
hyperfine states of the same atom.

\subsubsection{Phase diagram}%

\begin{figure}
[ptb]
\begin{center}
\includegraphics[
height=7.7936cm,
width=12.1166cm
]%
{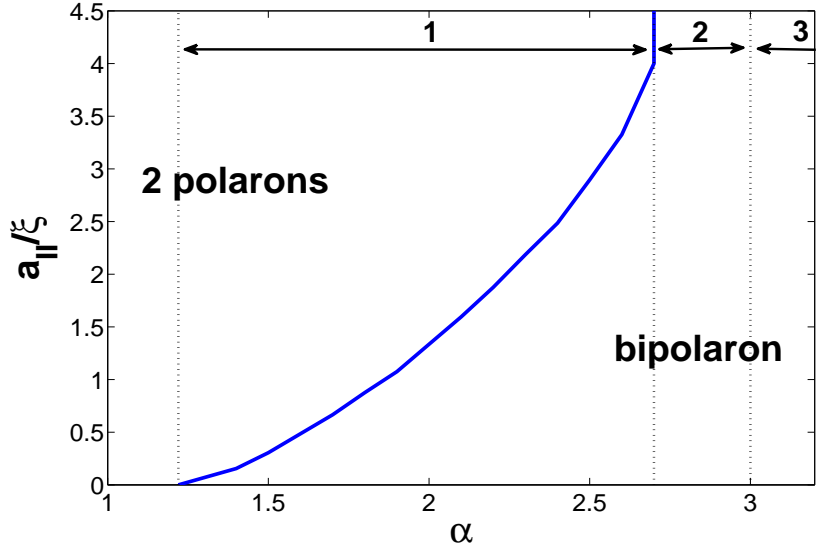}%
\caption{(Color online) The $\left(  a_{II},\alpha \right)  $-phase diagram for
two Li-6 impurties in a Na condensate with $\alpha$ the polaronic coupling
parameter and $a_{II}$ the impurity-impurity scattering length. The solid line
indicates the polaron-bipolaron transition. The dotted lines at $\alpha=1.22$,
$\alpha=2.71$ and $\alpha=3$ indicate the boundaries of the different regions,
as discussed in the text.}%
\label{Fig: PhaseDiag}%
\end{center}
\end{figure}
The upper bound for the ground state energy (\ref{JensFeynTemp0}) was
minimized as a function of the variational parameters $\left \{  \vec{a}%
,\Omega_{1},\Omega_{2},\Omega_{3},\nu \right \}  $ for given values of the
coupling parameter $\alpha$ and the impurity-impurity scattering length
$a_{II}$. If the resulting upper bound is lower than twice the upper bound for
the one-polaron ground state energy we conclude that it is energetically
favorable to form a bipolaron, otherwise the system consists of two separate
polarons. This procedure results in the $\left(  a_{II},\alpha \right)  $-phase
diagram presented in Fig. \ref{Fig: PhaseDiag} where we have also indicated
three regions as a function of $\alpha$. For $\alpha>2.71$ the formation of a
bipolaron is always energetically favorable, irrespective of $a_{II}$, and the
area with $\alpha \in \left[  2.71,3\right]  $ is denoted as region 2 (see Fig.
\ref{Fig: PhaseDiag}). At $\alpha=3$ the Feynman all-coupling single-polaron
treatment predicts the transition to the strong couping regime
\cite{PhysRevB.80.184504} and the area with $\alpha>3$ is denoted as region 3.
From Fig. \ref{Fig: PhaseDiag} it is clear that for $\alpha \in \left[
1.22,2.71\right]  $ (region 1) the bipolaron is only stable at small values of
$a_{II}$. For $\alpha<1.22$ a bipolaron is never formed. If the bipolaron is
stable the variationally determined vector $\vec{a}$, separating the
impurities in the trial system of Fig. \ref{fig: modelsysteem}, is always zero
in region 1, irrespective of $a_{II}$, while for $\alpha>2.71$ it is finite at
sufficiently large $a_{II}$ in which case the shape of the bipolaron can be
interpreted as a dumbbell.

\subsubsection{Bipolaron mass and radius}

\begin{figure}
[ptb]
\begin{center}
\includegraphics[
height=7.4575cm,
width=12.1166cm
]%
{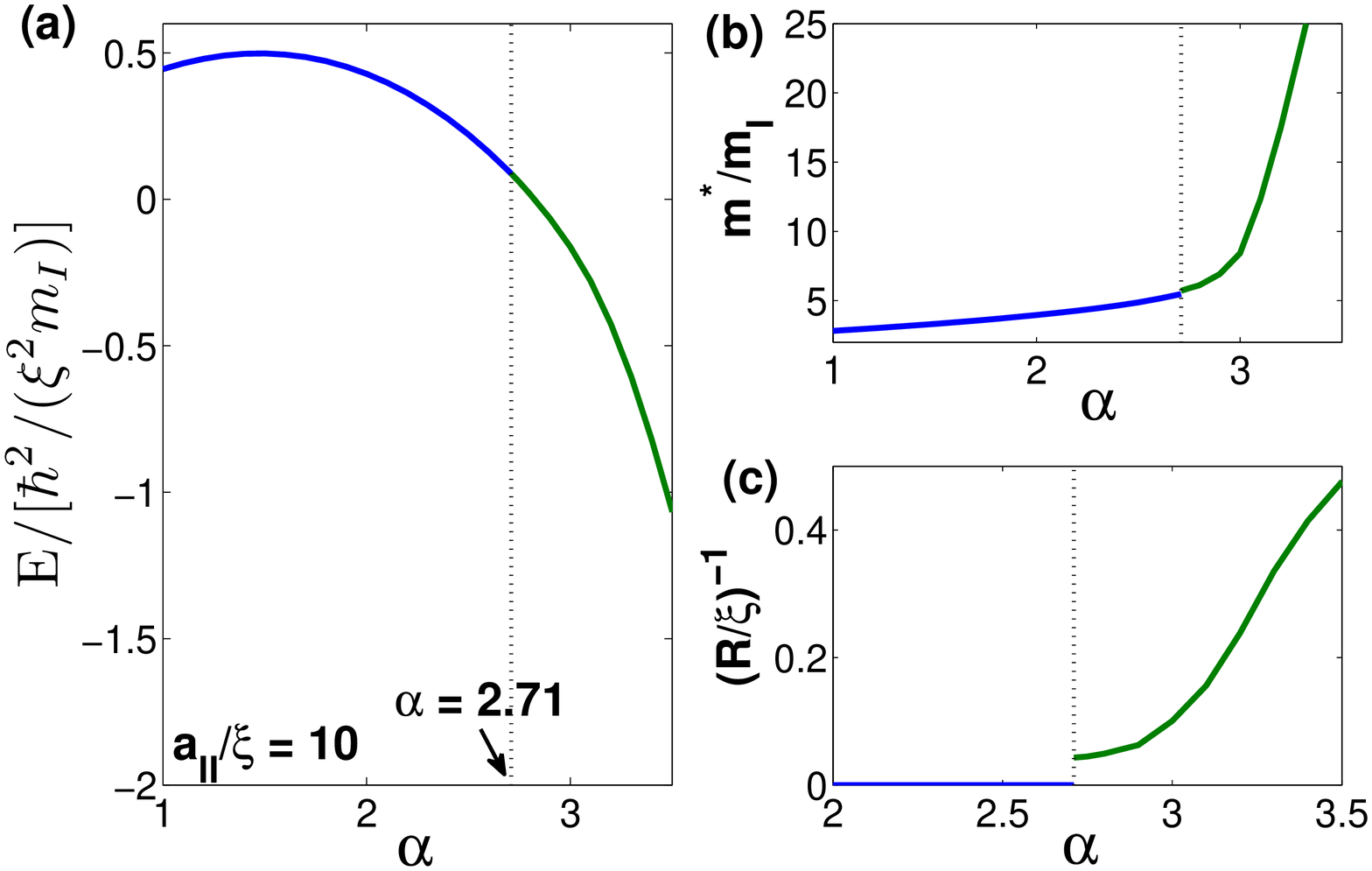}%
\caption{(Color online) The upper bound of the ground state energy (a), the
bipolaron effective mass (b) and the inverse of the bipolaron radius (c) are
presented as a function of the polaronic coupling parameter $\alpha$ at
$a_{II}=10$ for two Li-6 impurity in a Na condensate. The dotted line at
$\alpha=2.71$ indicates the polaron-bipolaron transition.}%
\label{Fig: aII10}%
\end{center}
\end{figure}

In Fig. \ref{Fig: aII10} the upper bound (\ref{JensFeynTemp0}), the polaronic
effective mass (\ref{BipEffMass}) and the inverse bipolaron radius
(\ref{BipRadius}) are presented as a function of the coupling parameter
$\alpha$ at $a_{II}=10$. For $\alpha<2.71$ the system consists of two separate
polarons while for $\alpha>2.71$ a bipolaron is formed. For two separate
polarons the bipolaron radius is defined as infinity. As a function of
$\alpha$ the effective mass exhibits an increasing behavior and for $\alpha>3$
it increases more rapidly, indicating the transition to the strong coupling
regime which is also present for a single polaron \cite{PhysRevB.80.184504}.
If the bipolaron is stable the bipolaron radius decreases as a function of
$\alpha$. This shows that the bipolaron becomes more tightly bound as the
coupling is increased.%

\begin{figure}
[ptb]
\begin{center}
\includegraphics[
height=7.7936cm,
width=12.1144cm
]%
{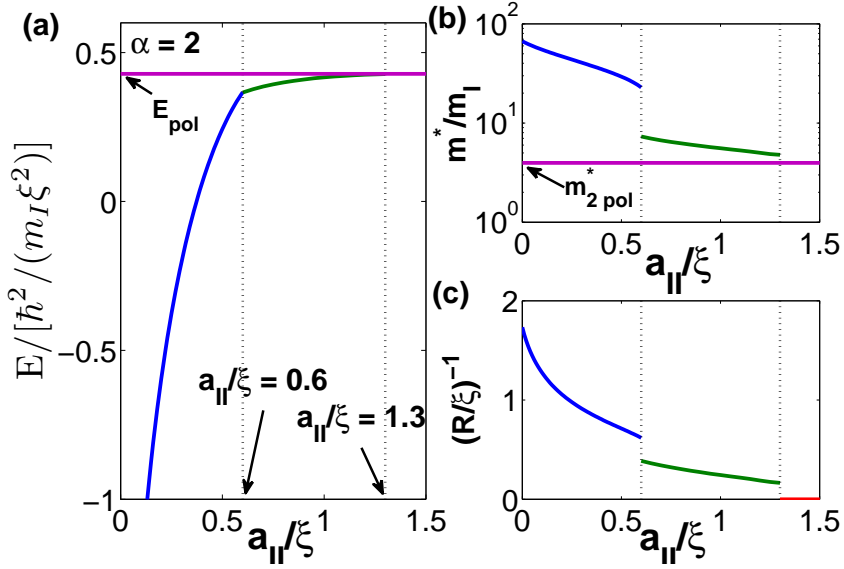}%
\caption{(Color online) The upper bound of the ground state energy (a), the
effective mass (b) (note the semi-logarithmic scale) and the inverse of the
bipolaron radius (c) are presented as a function of the impurity-impurity
scattering length $a_{II}$ at $\alpha=2$ (region 1 in Fig.
\ref{Fig: PhaseDiag}) for two Li-6 impurities in a Na condensate. The
polaron-bipolaron transition at $a_{II}/\xi=1.3$ and the transition of the
internal bipolaron state (see text) at $a_{II}/\xi=0.6$ are indicated with the
dotted lines. Also the upper bound for the ground state energy of two separate
polarons ($E_{2\text{ pol}}$) and the corresponding effective mass
($m_{2\text{ pol}}^{\ast}$) are shown.}%
\label{Fig: Alf2}%
\end{center}
\end{figure}

In Fig. \ref{Fig: Alf2} the upper bound (\ref{JensFeynTemp0}), the polaronic
effective mass (\ref{BipEffMass}) and the inverse bipolaron radius
(\ref{BipRadius}) are presented as a function of the impurity-impurity
scattering length $a_{II}$ at $\alpha=2$ (region 1 in Fig.
\ref{Fig: PhaseDiag}). The energy and the effective mass of two polarons are
also shown. This reveals that the bipolaron is only stable for sufficiently
small values of $a_{II}$, with a polaron-bipolaron transition at $a_{II}%
/\xi=1.3$. From Fig. \ref{Fig: Alf2} it is clear that in this case the
polaron-bipolaron transition is accompanied with a discontinuity in the
effective mass. The value $a_{II}/\xi=0.6$ is also indicated in Fig.
\ref{Fig: Alf2} which corresponds to another discontinuity in the effective
mass, as well as in the bipolaron radius. For $a_{II}/\xi<0.6$ the bipolaron
is relatively small and heavy and the ground state energy and the properties
exhibit a strong dependence on $a_{II}$ as compared to the behavior for
$a_{II}/\xi>0.6$. This suggests that for $a_{II}/\xi<0.6$ the bipolaron can be
considered as a tightly bound particle, while for $a_{II}/\xi>0.6$ it consists
of a more loosely bound state of two polarons. Increasing $a_{II}$ results in
a less tightly bound bipolaron and finally in the formation of two separate
polarons.%

\begin{figure}
[ptb]
\begin{center}
\includegraphics[
height=7.4575cm,
width=12.1166cm
]%
{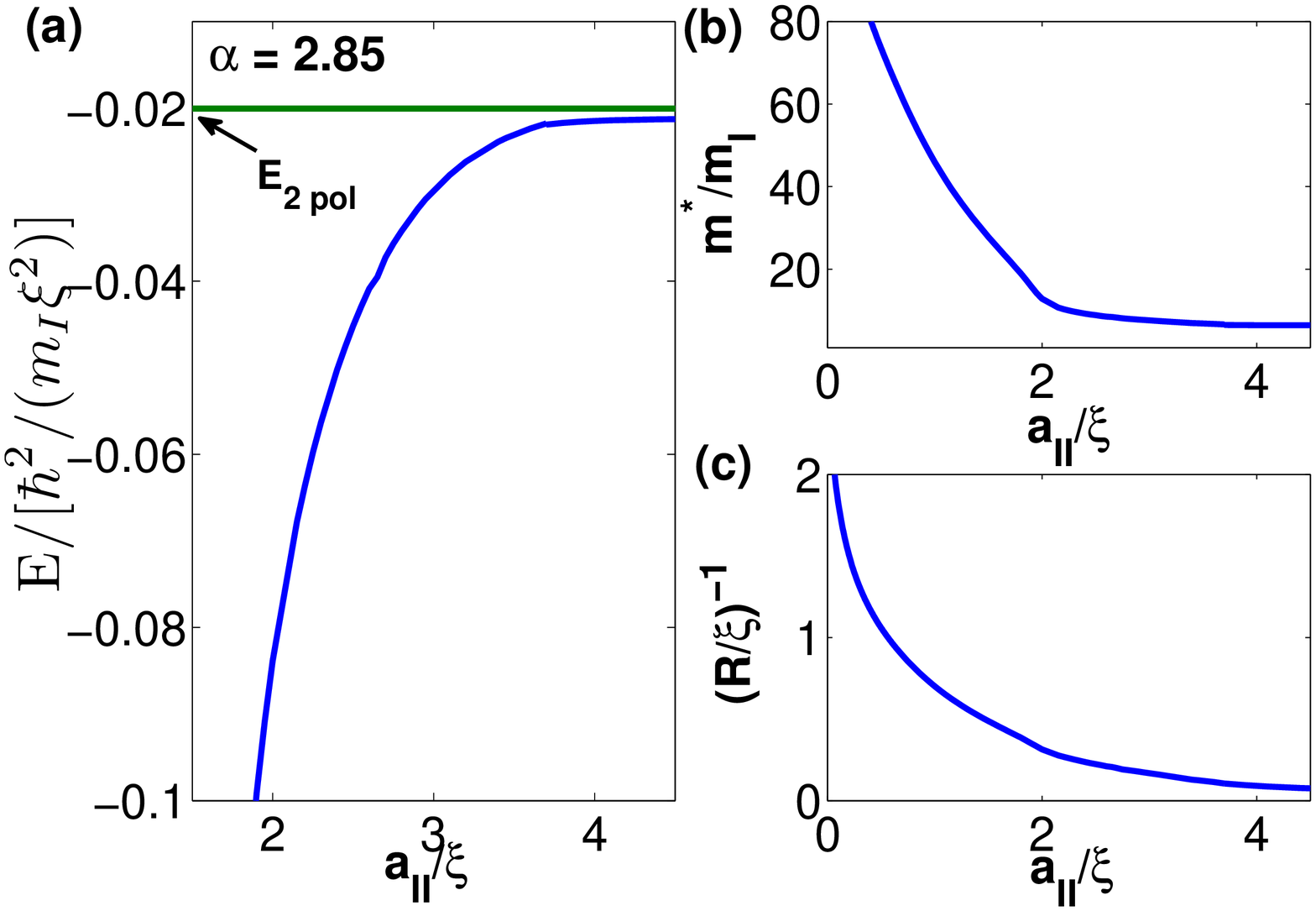}%
\caption{(Color online) The upper bound of the ground state energy (a), the
effective mass (b) and the inverse of the bipolaron radius (c) are presented
as a function of the impurity-impurity scattering length $a_{II}$ at
$\alpha=2.85$ (region 2 in Fig. \ref{Fig: PhaseDiag}) for two Li-6 impurities
in a Na condensate. The\ upper bound for the ground state energy of two
polarons ($E_{2\text{ pol}}$) is also indicated.}%
\label{fig: Alf285}%
\end{center}
\end{figure}

In Fig. \ref{fig: Alf285} the upper bound (\ref{JensFeynTemp0}), the polaronic
effective mass (\ref{BipEffMass}) and the inverse bipolaron radius
(\ref{BipRadius}) are presented as a function of the impurity-impurity
scattering length $a_{II}$ at $\alpha=2.85$ (region 2 in Fig.
\ref{Fig: PhaseDiag}). The energy of two separate polarons is also indicated
which shows that the bipolaron is always stable, irrespective of $a_{II}$.
Also here we find that increasing $a_{II}$ results in a less tightly bound
bipolaron. Again we can distinguish two regimes in the $a_{II}$-dependence of
$E$ and $m^{\ast}$, but now without a discontinuity at the transition. At
small $a_{II}$ the increase of the ground state energy and the decrease of the
effective mass as a function of $a_{II}$ are significantly faster than at
higher values of $a_{II}$, with a transition region at $a_{II}/\xi \approx2$.
As before, this indicates that at small values of $a_{II}$ the bipolaron can
be considered as a tightly bound particle while at large values it is more
appropriately interpreted as two loosely bound polarons. Moreover, at small
values for $a_{II}$ the variationally determined vector $\vec{a}$, separating
the two impurities in the trial system of Fig. \ref{fig: modelsysteem}, is
zero while for relatively large $a_{II}$ it is finite. This shows that the
loosely bound polarons at large $a_{II}$ are spatially separated and the shape
can be interpreted as a dumbbell.%

\begin{figure}
[ptb]
\begin{center}
\includegraphics[
height=8.8568cm,
width=12.1166cm
]%
{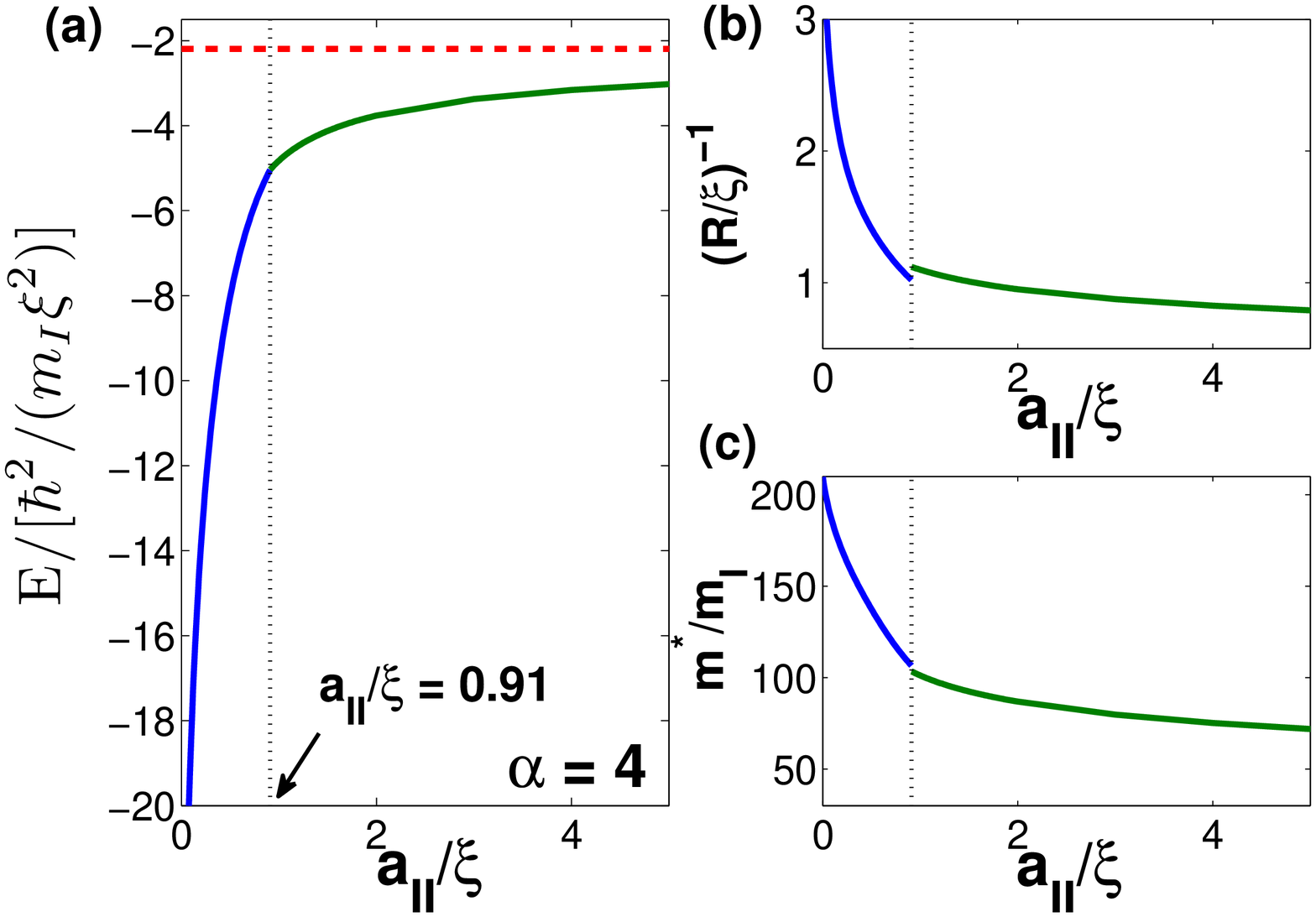}%
\caption{(Color online) The upper bound of the ground state energy (a), the
inverse of the bipolaron radius (b) and the effective mass (c) are presented
as a function of the impurity-impurity scattering length $a_{II}$ at
$\alpha=4$ (region 3 in Fig. \ref{Fig: PhaseDiag}) for two Li-6 impurities in
a Na condensate. The dotted line at $a_{II}=0.91\xi$ indicates a transition of
the internal state of the bipolaron (see text). The dashed line shows the
upper bound to the ground state energy in the limit $a_{II}\rightarrow \infty
$.}%
\label{Fig: Alf4}%
\end{center}
\end{figure}

Finally, in Fig. \ref{Fig: Alf4} \ the upper bound (\ref{JensFeynTemp0}), the
polaronic effective mass (\ref{BipEffMass}) and the inverse bipolaron radius
(\ref{BipRadius}) are presented as a function of the impurity-impurity
scattering length $a_{II}$ at $\alpha=4$ (region 3 in Fig.
\ref{Fig: PhaseDiag}). The upper bound (\ref{JensFeynTemp0}) in the limit
$a_{II}\rightarrow \infty$ is also indicated which is lower than the upper
bound of the ground state energy for two separate polarons. This shows that
the bipolaron is always stable. Also here we observe two regimes as a function
of $a_{II}$ with a transition at $a_{II}/\xi=0.91$ which is accompanied with a
discontinuity in the effective mass and the bipolaron radius, indicating a
transition between a tightly bound bipolaron and a more loosely bound state of
two polarons. Also in this case the variationally determined vector $\vec{a}$
is zero for the tightly bound bipolaron and non-zero for the loosely bound
state of two polarons, resulting in a dumbbell bipolaron.%

\section{Strong coupling formalism\label{section: SC}}

We now apply a generalization of the Landau-Pekar strong-coupling treatment
for $N_{I}$ polaronic particles to impurities in a Bose-Einstein condensate to
examine the formation of multi-polarons.

\subsection{Formalism\label{Sect: SCFormalism}}

For the description of the strong coupling formalism the product Ansatz is
used which states that the total wave function ($\left \vert \Phi \right \rangle
$) is the product of a part that describes the Bogoliubov excitations
($\left \vert \phi \right \rangle $) and a part for the impurities ($\left \vert
\Psi^{\left(  N_{I}\right)  }\right \rangle $): $\left \vert \Phi \right \rangle
=\left \vert \phi \right \rangle \left \vert \Psi^{\left(  N_{I}\right)
}\right \rangle $. Taking the expectation value of the Fr\"{o}hlich Hamiltonian
(\ref{HamNiPol}) with respect to $\left \vert \Phi \right \rangle $ and
completing the squares for the Bogoliubov creation and annihilation operators
results in:%
\begin{align}
\left \langle \Phi \left \vert \widehat{H}_{pol}^{\left(  N_{I}\right)
}\right \vert \Phi \right \rangle  &  =K+\sum_{\vec{k}\neq0}\hbar \omega_{\vec{k}%
}\left \langle \phi \left \vert \left(  \widehat{\alpha}_{\vec{k}}^{\dag}%
+\frac{V_{\vec{k}}\rho_{I}\left(  \vec{k}\right)  }{\hbar \omega_{\vec{k}}%
}\right)  \left(  \widehat{\alpha}_{\vec{k}}+\frac{V_{\vec{k}}^{\ast}\rho
_{I}^{\dag}\left(  \vec{k}\right)  }{\hbar \omega_{\vec{k}}}\right)
\right \vert \phi \right \rangle \nonumber \\
&  -\sum_{\vec{k}\neq0}\frac{\left \vert V_{\vec{k}}\widehat{\rho}_{I}\left(
\vec{k}\right)  \right \vert ^{2}}{\hbar \omega_{\vec{k}}}+U. \label{ExpHam}%
\end{align}
Where $K$ is the kinetic energy, $\rho_{I}\left(  \vec{k}\right)  $ the
Fourier transform of the density and $U$ the mutual interaction energy of the
impurities:%
\begin{align}
K  &  =\left \langle \Psi^{\left(  N_{I}\right)  }\left \vert \sum_{i=1}^{N_{I}%
}\frac{\widehat{p_{i}}^{2}}{2m_{I}}\right \vert \Psi^{\left(  N_{I}\right)
}\right \rangle ;\label{SCKin}\\
\rho_{I}\left(  \vec{k}\right)   &  =\left \langle \Psi^{\left(  N_{I}\right)
}\left \vert \sum_{i=1}^{N_{I}}e^{i\vec{k}.\widehat{\vec{r}}_{i}}\right \vert
\Psi^{\left(  N_{I}\right)  }\right \rangle ;\label{SCDens}\\
U  &  =\left \langle \Psi^{\left(  N_{I}\right)  }\left \vert \sum_{i<j}^{N_{I}%
}V_{II}\left(  \widehat{\vec{r}}_{i}-\widehat{\vec{r}}_{j}\right)  \right \vert
\Psi^{\left(  N_{I}\right)  }\right \rangle . \label{SCIntEn}%
\end{align}
The expectation value of the Hamiltonian (\ref{ExpHam}) is minimal if the wave
function of the Bogoliubov excitations is chosen as the vacuum $\left \vert
\phi_{g}\right \rangle $ for the "displaced" operators:%
\begin{equation}
\left \langle \phi_{g}\left \vert \left(  \widehat{\alpha}_{\vec{k}}^{\dag
}+\frac{V_{\vec{k}}\rho_{I}\left(  \vec{k}\right)  }{\hbar \omega_{\vec{k}}%
}\right)  \left(  \widehat{\alpha}_{\vec{k}}+\frac{V_{\vec{k}}^{\ast}\rho
_{I}^{\dag}\left(  \vec{k}\right)  }{\hbar \omega_{\vec{k}}}\right)
\right \vert \phi_{g}\right \rangle =0. \label{BogWavefun}%
\end{equation}
This results in the following expression for the ground state energy:%
\begin{equation}
E_{0}^{\left(  N_{I}\right)  }=K-\sum_{\vec{k}\neq0}\frac{\left \vert
V_{\vec{k}}\right \vert ^{2}\left \vert \rho_{I}\left(  \vec{k}\right)
\right \vert ^{2}}{\hbar \omega_{\vec{k}}}+U. \label{SCUpperBound}%
\end{equation}
This result can alternatively be derived with a canonical transformation, as
done by Bogolubov and Tyablikov for a single polaron \cite{BogTya}. For the
impurities we will use a variational wave function, the resulting energy
(\ref{SCUpperBound}) is then an upper bound for the ground state energy.

\subsubsection{Effective mass}

The strong-coupling formalism allows a derivation of the multi-polaron
effective mass in a similar way as was done for a single polaron in Refs.
\cite{Pekar, Bogol, Evrard1965295, CasteelsLaserPhys} and for the bipolaron in
Ref. \cite{0305-4470-27-23-035}. The total momentum of the polaronic system
$\widehat{\mathcal{\vec{P}}}$\ is given by%
\begin{equation}
\widehat{\mathcal{\vec{P}}}=\widehat{\vec{P}}+\sum_{\vec{k}}\hbar \vec
{k}\widehat{\alpha}_{\vec{k}}^{\dag}\widehat{\alpha}_{\vec{k}},
\end{equation}
with $\widehat{\vec{P}}=\sum_{i}^{N_{I}}\widehat{\vec{p}}_{i}$. This operator
commutes with the Fr\"{o}hlich Hamiltonian (\ref{HamNiPol}) and the total
momentum is thus a constant of motion: $\left \langle \widehat{\mathcal{\vec
{P}}}\right \rangle =\mathcal{\vec{P}}$. We make this explicit by means of a
Lagrange multiplier $\vec{v}$ which physically represents the velocity of the
system and consider the operator%
\begin{equation}
\widehat{H}_{pol}^{\left(  N_{I}\right)  }\left(  \vec{v}\right)  =\widehat
{H}_{pol}^{\left(  N_{I}\right)  }-\vec{v}.\left(  \widehat{\vec{P}}%
+\sum_{\vec{k}}\hbar \vec{k}\widehat{\alpha}_{\vec{k}}^{\dag}\widehat{\alpha
}_{\vec{k}}-\mathcal{\vec{P}}\right)
\end{equation}
for minimization. The effective mass $m^{\ast}$ of the multi-polaron can then
be determined from the relation $\mathcal{\vec{P}}=m^{\ast}\vec{v}$. The
impurity variational wave function $\left \vert \Psi^{\left(  N_{I}\right)
}\right \rangle $, with $\left \langle \widehat{\vec{P}}\right \rangle =0$, has
to be adapted to a wave function with finite averaged momentum $\left \langle
\widehat{\vec{P}}\right \rangle =N_{I}m_{I}\vec{v}$, we use%
\begin{equation}
\Psi^{\prime \left(  N_{I}\right)  }\left(  \left \{  \vec{r}_{i}\right \}
\right)  =\exp \left[  \frac{im_{I}\vec{v}.\sum \vec{r}_{i}}{\hbar}\right]
\Psi^{\left(  N_{I}\right)  }\left(  \left \{  \vec{r}_{i}\right \}  \right)  .
\end{equation}
Taking the expectation value of $\widehat{H}_{pol}^{\left(  N_{I}\right)
}\left(  \vec{v}\right)  $ with respect to the product wave function
$\left \vert \phi \right \rangle \left \vert \Psi^{\prime \left(  N_{I}\right)
}\right \rangle $ and introducing the wave function $\left \vert \phi
_{g}\right \rangle $ for the Bogoliubov excitations, as in (\ref{BogWavefun}),
results in:%
\begin{equation}
E^{\left(  N_{I}\right)  }\left(  \vec{v}\right)  =K+U-N_{I}\frac{m_{I}v^{2}%
}{2}+v.\mathcal{\vec{P}}-\sum_{\vec{k}\neq0}\frac{\left \vert V_{\vec{k}%
}\right \vert ^{2}\left \vert \rho_{I}\left(  \vec{k}\right)  \right \vert ^{2}%
}{\hbar \omega_{\vec{k}}-\hbar \vec{v}.\vec{k}}, \label{EnergyeFunV}%
\end{equation}
with $K$, $\rho_{I}\left(  \vec{k}\right)  $ and $U$ as defined in
(\ref{SCKin}), (\ref{SCDens}) and (\ref{SCIntEn}), respectively. Minimizing
expression (\ref{EnergyeFunV}) with respect to $\vec{v}$ and performing a
Taylor expansion for small $\vec{v}$ gives for the effective mass%
\begin{equation}
m^{\ast}=N_{I}m_{I}+2\hbar^{2}\sum_{\vec{k}\neq0}\frac{\left \vert V_{\vec{k}%
}\right \vert ^{2}\left \vert \rho_{I}\left(  \vec{k}\right)  \right \vert ^{2}%
}{\left(  \hbar \omega_{\vec{k}}\right)  ^{3}}k_{z}^{2}, \label{EffMass}%
\end{equation}
with $k_{z}$ the $z$-component of $\vec{k}$.

\subsubsection{Variational impurity wave function}

For $N_{I}$ distinguishable particles we consider the following normalized
variational wave function:%
\begin{equation}
\Psi^{\left(  N_{I}\right)  }\left(  \left \{  \vec{r}_{i}\right \}  \right)
=\prod_{i=1}^{N_{I}}\frac{1}{\left(  \pi \lambda^{2}\right)  ^{3/4}}\exp \left[
-\frac{\left(  \vec{r}_{i}-\vec{a}_{i}\right)  ^{2}}{2\lambda^{2}}\right]  ,
\label{VarWaveFunc}%
\end{equation}
which consists of $N_{I}$ Gaussians with standard deviation $\lambda$,
centered at $\vec{a}_{i}$. For a single impurity in a condensate a numerical
calculation of the wave function revealed a good agreement with a Gaussian
wave function if the polaronic coupling is strong enough
\cite{0295-5075-82-3-30004}. The corresponding expectation values are%
\begin{align}
K  &  =N_{I}\frac{3\hbar^{2}}{4m_{I}\lambda^{2}};\\
U  &  =\sum_{i<j}^{N_{I}}\sum_{\vec{k}}V_{II}\left(  \vec{k}\right)
e^{-\frac{k^{2}\lambda^{2}}{2}+i\vec{k}.\left(  \vec{a}_{1}-\vec{a}%
_{2}\right)  };\label{SCIntEn2}\\
\rho_{I}\left(  \vec{k}\right)   &  =\exp \left[  -\frac{k^{2}\lambda^{2}}%
{4}\right]  \sum_{i=1}^{N_{I}}e^{i\vec{k}.\vec{a}_{i}}.
\end{align}
The wave function (\ref{VarWaveFunc}) can be extended to the case of identical
impurities by using a Slater determinant for fermions or the appropriate
symmetrized wave function for bosons.

\subsubsection{Single polaron limit}

If all the impurities are infinitely separated we expect the multi-polaron to
reduce to individual polarons. For the impurity wave function
(\ref{VarWaveFunc}) this corresponds to the limit $\left \vert \vec{a}_{i}%
-\vec{a}_{j}\right \vert \rightarrow \infty$ $\forall i\neq j$. The\ upper bound
for the polaron ground state energy (\ref{SCUpperBound})\ becomes in this
limit:%
\begin{equation}
\lim_{\substack{\left \vert \vec{a}_{i}-\vec{a}_{j}\right \vert \rightarrow
\infty \\ \forall i\neq j}}E_{0}^{\left(  N_{I}\right)  }=N_{I}\left(
\frac{3\hbar^{2}}{4m_{I}\lambda^{2}}-\sum_{\vec{k}\neq0}\frac{\left \vert
V_{\vec{k}}\right \vert ^{2}}{\varepsilon_{\vec{k}}}e^{-\frac{k^{2}\lambda^{2}%
}{2}}\right)  =N_{I}E_{0}^{\left(  1\right)  }.
\end{equation}
This equals $N_{I}$ times the strong-coupling result for the upper bound for
the ground state energy of a single polaron ($E_{0}^{\left(  1\right)  }$), as
expected. The effective mass of the multi-polaron (\ref{EffMass}) becomes in
this limit:%
\begin{equation}
\lim_{\substack{\left \vert \vec{a}_{i}-\vec{a}_{j}\right \vert \rightarrow
\infty \\ \forall i\neq j}}m^{\ast}=N_{I}\left(  m_{I}+2\hbar^{2}\sum_{\vec
{k}\neq0}\frac{\left \vert V_{\vec{k}}\right \vert ^{2}}{\left(  \hbar
\omega_{\vec{k}}\right)  ^{3}}e^{-\frac{k^{2}\lambda^{2}}{2}}k_{z}^{2}\right)
=N_{I}m_{pol}^{\ast}.
\end{equation}
This is $N_{I}$ times the strong-coupling result for the effective mass of a
single polaron ($m_{pol}^{\ast}$), again as expected.

\subsubsection{Impurities in a Bose-Einstein condensate}

We now consider the specific polaronic system consisting of impurities in a
BEC. Introducing the Bogoliubov dispersion (\ref{BogDisp}), the interaction
amplitude (\ref{IntAmp}) and the variational wave function (\ref{VarWaveFunc})
for $N_{I}$ distinguishable impurities in the upper bound for the ground state
energy (\ref{SCUpperBound}) gives in polaronic units ($\hbar=m_{I}=\xi=1$):
\begin{align}
E_{0}^{\left(  N_{I}\right)  }  &  =\frac{3N_{I}}{4\lambda^{2}}+\frac{2a_{II}%
}{\left(  2\pi \right)  ^{1/2}\lambda^{3}}\sum_{i<j}^{N_{I}}\exp \left[
-\frac{\left(  \vec{a}_{i}-\vec{a}_{j}\right)  ^{2}}{2\lambda^{2}}\right]
\nonumber \\
&  -\frac{2\alpha \mu}{\pi}\int_{0}^{\infty}dkk^{2}\frac{\exp \left[
-\frac{k^{2}\lambda^{2}}{2}\right]  }{k^{2}+2}\sum_{i,j}^{N_{I}}\frac
{\sin \left[  k\left \vert \vec{a}_{i}-\vec{a}_{j}\right \vert \right]
}{k\left \vert \vec{a}_{i}-\vec{a}_{j}\right \vert }, \label{SCUpperBoundImps}%
\end{align}
where a contact pseudo potential with the first order Lippmann-Schwinger
result was used for the impurity-impurity interaction. We also introduced
expression (\ref{CouplPar}) for the dimensionless polaronic coupling parameter
$\alpha$ and the dimensionless mass factor%
\begin{equation}
\mu=\frac{\left(  m_{B}+m_{I}\right)  ^{2}}{4m_{B}m_{I}}.
\end{equation}
In the strong coupling regime the mass parameter $m_{B}/m_{I}$ and the
coupling parameter $\alpha$ combine to a single dimensionless coupling
parameter $\alpha \mu$. The effective mass of $N_{I}$ impurities in a BEC in
polaronic units can be written as%
\begin{equation}
m^{\ast}=N_{I}m_{I}+\frac{4\alpha \mu m_{B}^{2}}{\pi^{2}}\int d\vec{k}%
\frac{\left \vert \rho_{I}\left(  \vec{k}\right)  \right \vert ^{2}}{\left(
2+k^{2}\right)  ^{2}}\frac{k_{z}^{2}}{k^{2}}. \label{SCEffMassImps}%
\end{equation}

\subsection{Results}

\subsubsection{Bipolaron}

First, we examine two impurities in a Bose-Einstein condensate ($N_{I}=2$) and
the formation of a bipolaron. The bipolaron radius, estimated by the mean
square distance between the impurities (\ref{BipRadius}), gives for the
impurity variational wave function (\ref{VarWaveFunc}):%
\begin{equation}
R=\sqrt{\left \langle \left(  \vec{r}_{1}-\vec{r}_{2}\right)  ^{2}\right \rangle
}=\sqrt{a^{2}+3\lambda^{2}}, \label{SCBipRad}%
\end{equation}
with $a=\left \vert \vec{a}_{1}-\vec{a}_{2}\right \vert $.%

\begin{figure}
[ptb]
\begin{center}
\includegraphics[
height=7.7936cm,
width=12.1144cm
]%
{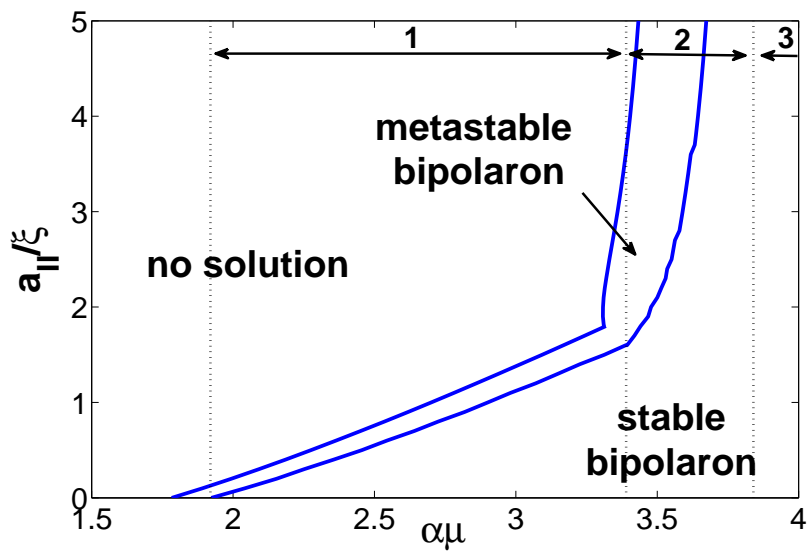}%
\caption{(Color online) The solid lines show the $\left(  a_{II},\alpha
\mu \right)  $-phase diagram for two distinguishable impurity atoms in a
Bose-Einstein condensate with $\alpha \mu$ the coupling parameter and $a_{II}$
the impurity-impurity scattering length as calculated with the strong-coupling
formalism. At strong coupling the bipolaron is stable and at sufficiently weak
coupling there is no solution. In between there is a region where the solution
results in a positive ground state energy, which means the bipolaron is
metastable. The dotted lines at $\alpha \mu=1.92$, $\alpha \mu=3.39$ and
$\alpha \mu=3.84$ indicate the boundaries of the different regions, as
discussed in the text.}%
\label{Fig: FaseDiagDis}%
\end{center}
\end{figure}
In Fig. \ref{Fig: FaseDiagDis} the $\left(  a_{II},\alpha \mu \right)  $-phase
diagram of two distinguishable impurity atoms in a BEC is presented and the
different regions as a function of $\alpha$ are also indicated. With each
$a_{II}$ a value $\left(  \alpha \mu \right)  _{exist}$ is associated in such a
way that for $\alpha \mu<\left(  \alpha \mu \right)  _{exist}$ the minimization
of the right-hand side of expression (\ref{SCUpperBoundImps}) yields no
solution at a finite value of $\lambda$. If $\alpha \mu>\left(  \alpha
\mu \right)  _{exist}$ another value $\left(  \alpha \mu \right)  _{stable}$ can
be determined, separating a region with a metastable bipolaron (positive
ground state energy) for $\alpha \mu<\left(  \alpha \mu \right)  _{stable}$ from
a region with a stable bipolaron (negative ground state energy) for $\alpha
\mu>\left(  \alpha \mu \right)  _{stable}$. In the case of a metastable
bipolaron the impuritites are expected to be expelled from the condensate.
However, since the formalism is only expected to be valid at strong coupling,
the physical relevance of the metastable bipolaron is not obvious. In the
limit $a_{II}\rightarrow \infty$ the strong-coupling results for a single
polaron are retrieved: $\left(  \alpha \mu \right)  _{exist}=3.57$ and $\left(
\alpha \mu \right)  _{stable}=3.84$ \cite{CasteelsLaserPhys}. For $\alpha
\mu>3.84$ (region 3) the formation of the bipolaron is always energetically
favorable as compared to the formation of two separate polarons, irrespective
of $a_{II}$. For $\alpha \mu \in \left[  3.39,3.84\right]  $ (region 2) a stable
solution only exists if $a_{II}$ is smaller than a critical value which is
relatively large and increases rapidly as a function of $\alpha \mu$. For
$\alpha \mu \in \left[  1.92,3.39\right]  $ (region 1) a stable bipolaron also
only exists for $a_{II}$ smaller than a critical value, but now this critical
value is relatively low and increases much more slowly as a function of
$\alpha \mu$, as compared to the behavior in region 2. At weaker coupling there
is never a stable solution. Considering the stable solution, the variationally
determined vector $\vec{a}_{1}-\vec{a}_{2}$, which represents the separation
between the two Gaussians in the wave function (\ref{VarWaveFunc}), is always
zero if $\alpha \mu<3.39$, for $\alpha \mu>3.39$ it is finite at sufficiently
large values of $a_{II}$, resulting in a dumbbell bipolaron.%
\begin{figure}
[ptb]
\begin{center}
\includegraphics[
height=8.8568cm,
width=12.1166cm
]%
{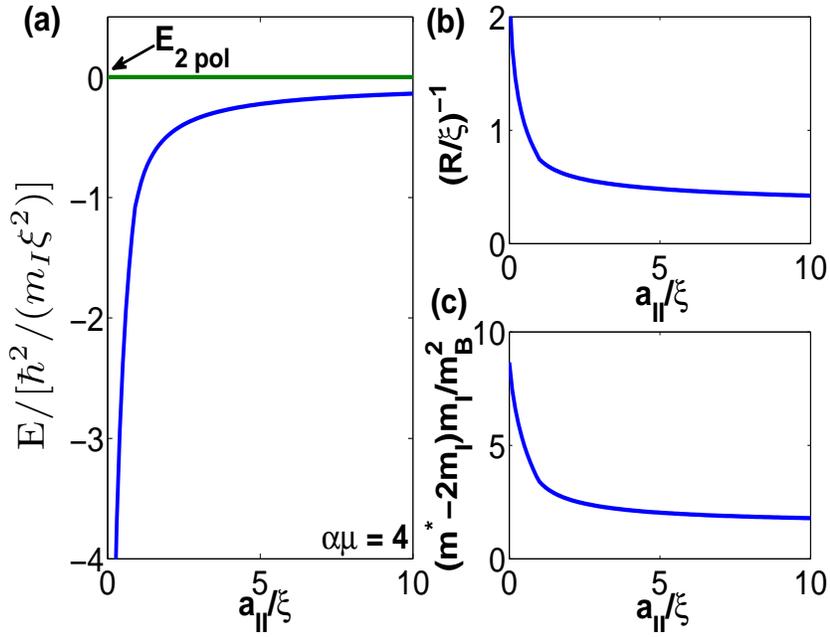}%
\caption{(Color online) The upper bound for the ground state energy (a), the
inverse bipolaron radius (b) and the effective mass (c) as a function of the
impurity-impurity scattering length $a_{II}$ at $\alpha \mu=4$ for two
distinguishable impurities in a Bose-Einstein condensate. The upper bound for
the ground state energy of two polarons ($E_{2\text{ pol}}$) is also
indicated.}%
\label{Fig: Alf4Dis}%
\end{center}
\end{figure}

In Fig. \ref{Fig: Alf4Dis} the upper bound for the ground state energy
(\ref{SCUpperBoundImps}), the inverse bipolaron radius (\ref{SCBipRad}) and
the effective mass (\ref{SCEffMassImps}) are presented as a function of the
impurity-impurity scattering length $a_{II}$ for two\ distinguishable
impurities in a BEC at $\alpha \mu=4$. The upper bound for the ground state
energy of a single polaron is also indicated which shows that the formation of
a bipolaron is energetically favorable for all finite values of $a_{II}$. From
Fig. \ref{Fig: Alf4Dis} we see that for increasing $a_{II}$ the bipolaron
binding energy decreases, the radius increases and the effective mass
decreases, showing that the bipolaron becomes less tightly bound. In the limit
$a_{II}\rightarrow \infty$ the bipolaron effective mass becomes twice the
effective mass of a single polaron. Considering the $a_{II}$-dependence of the
properties reveals two regimes with a transtition at $a_{II}/\xi=0.85$. For
$a_{II}/\xi<0.85$ there is a relatively strong dependence of the properties on
$a_{II}$ as compared to the behavior for $a_{II}/\xi>0.85$. Furthermore, the
variationally determined vector $\vec{a}_{1}-\vec{a}_{2}$, representing the
distance between the two centers of the Gaussians in the wave function
(\ref{VarWaveFunc}), is only non-zero for $a_{II}/\xi>0.85$. This is
consistent with our earlier interpretations, in that for $a_{II}/\xi<0.85$ the
bipolaron is tightly bound while for $a_{II}/\xi>0.85$ it is a more loosely
bound dumbbell bipolaron.

A similar analysis can be made for identical impurities by anti-symmetrizing
the wave function (\ref{VarWaveFunc}) for fermions or symmetrizing it for
bosons. In the case of two identical bosons the same qualitative results are
retrieved as for distinguishable impurities. For identical fermions the
symmetry of the wave function results in a vanishing expectation value of the
s-wave contact pseudopotential (\ref{SCIntEn2}) which implies that at ultra
low temperatures identical fermions behave as non-interacting particles. The
only remaining parameter is the coupling parameter $\alpha \mu$ and we find
$\left(  \alpha \mu \right)  _{exist}=3.09$ as a minimum for a solution to exist
and $\left(  \alpha \mu \right)  _{stable}=3.31$ as a minimum to find a stable
solution. Furthermore, if a solution exists, it is always energetically
favorable to form a bipolaron as compared to two separate polarons.

\subsubsection{Multi-polaron}

We now examine the multi-polaron by minimizing the upper bound for the ground
state energy (\ref{SCUpperBoundImps}) for $N_{I}=1,2,...,8$ distinguishable
impurities. In Ref. \cite{1367-2630-13-10-103029} a similar procedure was
presented, but the expectation values of the positions of the impurities was
considered equal which corresponds to the variational wave function
(\ref{VarWaveFunc}) with $\vec{a}_{i}=\vec{a}_{j}$ $\forall i\neq j$.

\begin{figure}
[ptb]
\begin{center}
\includegraphics[
height=11.3038cm,
width=12.1144cm
]%
{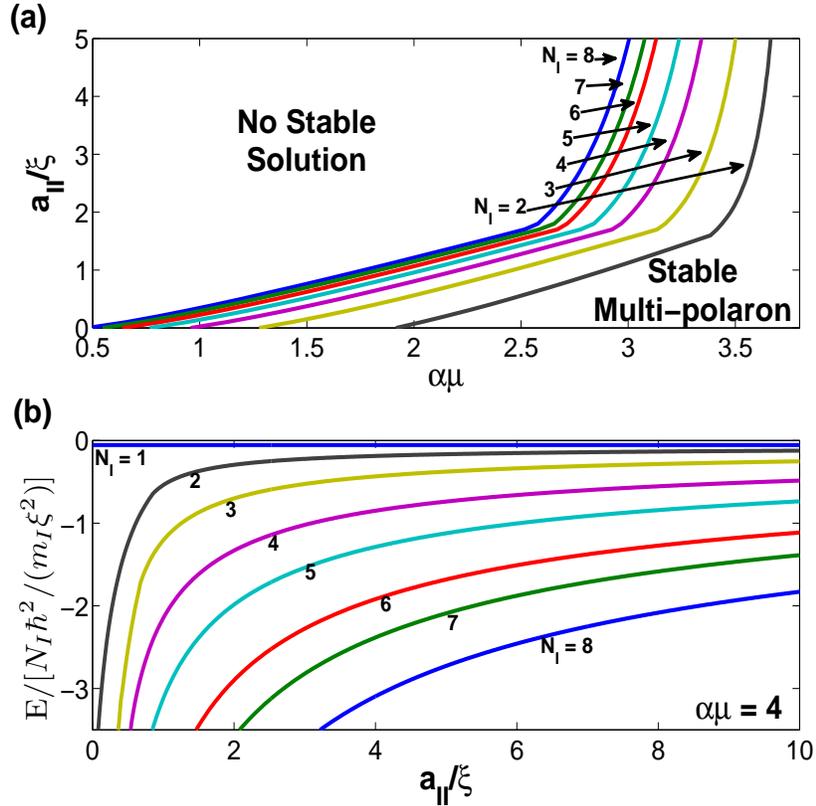}%
\caption{(Color online) In (a) the $\left(  a_{II},\alpha \mu \right)  $-phase
diagram for $N_{I}$ distinguishable impurities in a Bose-Eintein condensate
with respect to the formation of a stable multi-polaron is presented with
$\alpha \mu$ the coupling parameter and $a_{II}$ the impurity-impurity
scattering length. In (b) the upper bound for the multi-polaron ground state
energy per impurity is presented as a function of the impurity-impurity
scattering length $a_{II}$ for distinguishable impurities at $\alpha \mu=4$. }%
\label{Fig: Multi}%
\end{center}
\end{figure}

In Fig. \ref{Fig: Multi} (a) the resulting phase diagram is presented for
$N_{I}=1,2,...,8$ distinguishable impurities in a Bose-Einstein condensate as
a function of the polaronic coupling $\alpha \mu$ and the impurity-impurity
scattering length $a_{II}$. Similar to before for the bipolaron for each
$a_{II}$ a minimum coupling value $\left(  \alpha \mu \right)  _{exist}$ is
associated for a solution of the minimization of (\ref{SCUpperBoundImps}) to
exist at finite $\lambda$ and another minimum value $\left(  \alpha \mu \right)
_{stable}$ ($\left(  \alpha \mu \right)  _{stable}>\left(  \alpha \mu \right)
_{exist}$) to find a stable solution. In the limit $a_{II}\rightarrow \infty$
the one-polaron strong-coupling results are found for any number of
impurities: $\left(  \alpha \mu \right)  _{exist}=3.57$ and $\left(  \alpha
\mu \right)  _{stable}=3.84$. In Fig. \ref{Fig: Multi} (a) only $\left(
\alpha \mu \right)  _{stable}$ is presented for clarity. This shows that if the
number of impurities $N_{I}$ is increased $\left(  \alpha \mu \right)
_{stable}$ decreases, resulting in a larger stability region. This behavior of
a smaller critical coupling value for the formation of a larger multi-polaron
was also observed in Ref. \cite{1367-2630-13-10-103029}.%

The $\left(  a_{II},\alpha \mu \right)  $-phase diagram in Fig. \ref{Fig: Multi}
(a) for a specific number $N_{I}$ of impurities is qualitatively the same as
for the bipolaron in Fig. \ref{Fig: FaseDiagDis}. This means the same
qualitative regions can be distinguished as a function of $\alpha \mu$ as we
did for the bipolaron in Fig. \ref{Fig: FaseDiagDis} and we present a general
analysis, valid for every value of $N_{I}$. In region 3 the formation of the
multi-polaron is always energetically favored as compared to $N_{I}$ separate
polarons. Furthermore, the variationally determined locations of the
impurities $\vec{a}_{i}$ coincide at sufficiently small $a_{II}$, indicating a
tightly bound multi-polaron, while at sufficiently large $a_{II}$ we find
$\vec{a}_{i}\neq \vec{a}_{j}$ $\forall i\neq j$, indicating a bound droplet of
$N_{I}$ separate polarons (cfr. the dumbbell bipolaron). In region 2 the
minimization of the right-hand side of (\ref{SCUpperBoundImps}) yields no
stable solution if $a_{II}$ is larger than a critical value which is
relatively large and increases rapidly as a function of $\alpha \mu$. In region
1 there is also no stable solution for $a_{II}$ above a critical value which
is now relatively low and increases slowly as a function of $\alpha \mu$. In
this regime the behavior of the critical $a_{II}$ as a function of $\alpha \mu$
resembles a straight line and it was shown in Ref.
\cite{1367-2630-13-10-103029} that in the limit $N_{I}\rightarrow \infty$ this
line is well-approximated by the boundary for phase separation
\cite{FaseScheiding, PhysRevLett.82.2228, PhysRevA.58.4836,
PhysRevLett.81.5718}:%
\begin{equation}
\frac{a_{II}}{\xi}=\alpha \mu \text{.}%
\end{equation}
At small $\alpha \mu$ there is never a solution.

In Fig. \ref{Fig: Multi} (b) the upper bound for the multi-polaron ground
state energy per impurity is presented as a function of the impurity-impurity
scattering length $a_{II}$ at $\alpha \mu=4$. For $a_{II}\rightarrow \infty$ all
curves tend to the single-polaron result. At finite $a_{II}$ the ground state
energy per particle decreases as $N_{I}$ is increased, showing that it is
energetically favorable for the impurities to cluster and form a multi-polaron.

For positive impurity-boson scattering length there is a depletion of the
condensate in the vicinity of the multi-polaron. This can be detrimental for
the polaronic description since the Bogoliubov approximation breaks down at
large depletion. This restriction is only important for the tightly bound
multi-polaron at small $a_{II}$. In the case of a loosely bound droplet of
separate polarons at sufficiently large $a_{II}$ the mean distance between the
impurities is of the order of the healing length which shows that the
depletion is spread out over a large volume and the Bogoliubov approximation
is not jeopardized. For the tightly bound multi-polaron on the other hand
these considerations result in a critical number of impurities above which the
system can not be described by the Fr\"{o}hlich Hamiltonian. It was shown in
Ref. \cite{1367-2630-13-10-103029} that for typical experimental parameters
only tightly bound multi-polarons with a few impurities are possible.

\section{Comparison of the bipolaron results from the two
formalisms\label{section: comp}}

The results of the strong-coupling formalism of section
\ref{Sect: SCFormalism}, applied to the specific system of two distinguishable
Li-6 impurities in a Na condensate ($m_{B}/m_{I}=3.82$ and $\mu=1.52$) can be
compared with the results of the path integral treatment of Sec.
\ref{section: JenFey}. The phase diagrams in Figs. \ref{Fig: PhaseDiag} and
\ref{Fig: FaseDiagDis} exhibit a similar qualitative behavior and we can
compare the three denoted regions in the figures separately. Both formalisms
predict that there is no formation of a bipolaron if the polaronic coupling is
to weak and at slightly stronger coupling (region 1) the bipolaron is only
formed at relatively small $a_{II}$. In region 2 the all-coupling approach
predicts always formation of the bipolaron while according to the
strong-coupling approach there is no bipolaron formation at very high values
of $a_{II}$. In region 3 both formalisms agree on the prediction that the
bipolaron is always formed, irrespective of $a_{II}$. Quantitatively the
strong-coupling formalism underestimates the critical coupling parameter for
bipolaron formation, as compared to the all-coupling approach. Note that the
strong-coupling approach also underestimates the critical coupling value for
the transition to the strong coupling regime, as compared to a numerical study
\cite{0295-5075-82-3-30004}.

Considering the $a_{II}$-dependence of the properties both formalisms reveal
two distinct regimes. The behavior at relatively small $a_{II}$ corresponds to
a tightly bound bipolaron that behaves as a single-particle while at
relatively large $a_{II}$ it is better interpreted as a loosely bound dumbbell
bipolaron. The all-coupling approach predicts a possible discontinuity in the
polaronic properties at this transition which diminishes as the coupling is
increased and ultimately, well in the strong-coupling regime, vanishes, as
also predicted by the strong-coupling treatment.

\section{Conclusions\label{section: concl}}

The Feynman all-coupling polaron treatment was applied for two distinguishable
impurities in a condensate. This showed that if the polaronic coupling is
strong enough a bipolaron is formed. We also calculated the bipolaron
effective mass and the bipolaron radius. Considering the dependence of the
polaronic properties on the impurity-impurity scattering length $a_{II}$
results in the distinction of two regimes as a function of $a_{II}$, a tightly
bound bipolaron that behaves as a single particle at relatively small $a_{II}$
and a more loosely bound dumbbell bipolaron at sufficiently large $a_{II}$. If
the coupling is sufficiently strong or weak this transition is found to be
accompanied by a discontinuity in the properties of the bipolaron which
becomes less pronounced as the coupling is increased towards the
strong-coupling regime.

We also applied a strong-coupling treatment to impurities in a Bose-Einstein
condensate. For two distinguishable impurities in a BEC this leads to similar
results as found by the all-coupling treatment. This strong-coupling treatment
was then extended for identical impurities. For identical bosons this results
in the same qualitative results as for distinguishable impurities. For
identical fermions the mutual s-wave interaction vanishes and above a critical
coupling strength the formation of a bipolaron is always energetically favored
as compared to two separate polarons.

The strong-coupling treatment was then applied for more impurities in a BEC to
consider the formation of multi-polarons. We find that the multi-polaron
becomes stable at weaker coupling as the number of impurities is increased.
Furthermore, the ground state energy per particle decreases as $N_{I}$ is
increased which shows that clustering is energetically favorable in the strong
coupling regime.

Since both formalisms are variational and depend on the choice of a model
system there is no guarantee that they describe the actual system. However,
the extension of the all-coupling Feynman approach for the bipolaron not only
uses the model system to describe the system but also incorporates corrections
and is expected to be more accurate as compared to the usual variational
principle. Furthermore, the agreement between the two formalisms at strong
coupling indicates that physics of the system is captured by the models.

\bibliographystyle{elsart-num}
\bibliography{bipolaron}

\end{document}